# Can the Tully-Fisher Relation be the Result of Initial Conditions?

Daniel J. Eisenstein[1] and Abraham Loeb
Astronomy Department, Harvard University,
60 Garden St., Cambridge MA 02138

## ABSTRACT

We use Monte Carlo realizations of halo formation histories and a spherical accretion model to calculate the expected scatter in the velocity dispersions of galactic halos of a given mass due to differences in their formation times. Assuming that the rotational velocity of a spiral galaxy is determined by the velocity dispersion of its halo and that its luminosity is related to its total baryonic mass, this scatter translates to a minimum intrinsic scatter in the Tully-Fisher relation. For popular cosmological models we find that the scatter due to variations in formation histories is by itself greater than allowed by observations. Unless halos of spiral galaxies formed at high redshift ($z \gtrsim 1$) and did not later accrete any significant amount of mass, the Tully-Fisher relation is not likely to be the direct result of cosmological initial conditions but rather a consequence of a subsequent feedback process.

*Subject headings:* cosmology: theory–galaxies: formation–galaxies: distances and redshifts

---

[1]Also at: Physics Department, Harvard University

## 1. Introduction

One of the most surprising observational results in extragalactic astronomy is that the luminosity $L$ and rotational velocity $v_c$ of a spiral galaxy obey a tight power-law relation, $L \propto v_c^\alpha$, with $\alpha$ ranging between a value of 2 in the blue and a value of 4 in the infrared (Willick & Strauss 1995, and references therein). Because $v_c$ can be directly measured while the inferred value for $L$ depends on distance, this so-called Tully-Fisher relation (Tully & Fisher 1977; Aaronson, Huchra, & Mould 1979) has been widely used as a tool to measure distances. By now, various catalogs have been compiled with an interest in calibrating and applying this relation. Differences in observational technique, such as bandpass, the method for observing the rotational velocity, and sample selection, affect the power-law slope and degree of scatter in the relation, with I band observations providing the tightest fit and $\alpha \approx 2.7$ (Mathewson et al. 1992). In the largest samples the scatter in the relation is around 0.4 mag over two decades in luminosity (Mathewson & Ford 1994; Willick et al. 1995a,b,c), while in other samples, scatter as low as 0.1 mag has been reported (Bernstein et al. 1994). Because of the correspondingly low distance errors, the Tully-Fisher relation has been widely used in peculiar velocity studies (Strauss & Willick 1995 and references therein).

Despite the popularity of the relation, the reason behind its existence is still unknown. It is particularly surprising that the complex processes of gravitational collapse and star formation do not act to blur this tight relation. In this paper we address the most basic question in this context, namely: *Could the low scatter observed in the Tully-Fisher relation be the direct result of cosmological initial conditions?* We therefore wish to relate the two quantities in the relation, the circular velocity and the luminosity, to properties of the dark matter halos in which the spiral galaxies form. The former depends on the shape and depth of the halo potential-well, while the latter depends on the total baryonic mass. If the cosmological scatter in these properties is greater than allowed by observations, then the Tully-Fisher relation would imply a subsequent feedback process that regularizes gas dynamics and star formation in galaxies according to the depth of their potential wells.

In this paper we study the spread in the potential depth of an ensemble of galaxy halos of a given mass due to the differences in their formation times. We construct Monte Carlo realizations of the formation histories of halos as provided by the excursion set formalism (Bond et al. 1991, hereafter BCEK; Lacey & Cole 1993, 1994, hereafter LCa,b). The history of a given halo is used to estimate its current binding energy and velocity dispersion under the most regularizing assumption that the halo relaxes to an isothermal configuration through spherical accretion. This method and its minimum scatter estimates are described in §2 for different halo masses and cosmological models. In §3 we relate these results to the observed scatter in the Tully-Fisher relation. We conclude that the implied minimum scatter due to the variations in halo formation histories is too high to be reconciled with observations.

## 2. Method and Results

The universality of flat rotation curves and the Tully-Fisher relation imply that spiral galaxies obey a regular behaviour. In order to show that this regularity cannot naturally emerge from a random set of initial conditions, it is sufficient to consider the most optimistic model in which all galaxy halos form as isothermal spheres. Variations on this assumption would only tend to smear further any regularity we find. Collisionless N-body simulations and analytic results indicate that halos do indeed tend toward distributions resembling isothermal spheres (Dubinski & Carlberg 1991; Katz & Gunn 1991; Spergel & Hernquist 1992).

The potential depth of an isothermal halo is simply characterized by its one-dimensional velocity dispersion $\sigma$. If all, or some fixed fraction, of the baryons in the halo cool to the galactic disk, then a perfect relation between the velocity dispersion $\sigma$ and the mass of the halo $M$ would produce virtually no scatter in the Tully-Fisher relation. This follows from the most optimistic assumption that the circular velocity of the disk at large radius is simply proportional to the velocity dispersion of the halo, neglecting any complications due to triaxiality of the halo, ellipticity of the galactic disk (Franx & de Zeeuw 1992), or variations in the rotation curve due to self-gravity of the baryonic disk.

However, as illustrated by the spherical collapse model (Gunn & Gott 1972; White 1995), $\sigma$ is not simply a function of $M$. Because $\sigma^2$ characterizes the kinetic energy of a virialized system, we can fix $\sigma$ by requiring that energy be conserved through the virialization process. For the collapse of a spherical shell without a cosmological constant ($\Lambda = 0$), the shell energy per unit mass is $dE/dM = -GM/R_{ta}$, where $M$ is the mass enclosed within the shell and $R_{ta}$ is the radius of the shell at turn-around. For $\Lambda = 0$ cosmologies, the spherical shell solution gives

$$R_{ta} \propto M^{1/3} t_{\text{coll}}^{2/3}, \quad (1)$$

where $t_{\text{coll}}$ is the time of collapse. We may then integrate over all shells to find the total energy $E$, which yields the velocity dispersion $\sigma^2 \propto E/M$. If an object collapses at a single time, $\sigma \propto M^{1/3} t_{\text{coll}}^{-1/3}$. The constant of proportionality is unimportant for this work.

During the hierarchical process of structure formation, the formation of an object is not characterized by a single redshift of collapse. Rather, the object is assembled over time by a combination of mergers and smooth accretion. Thus, one would expect variations in these formation histories to produce a scatter in the velocity dispersion for an ensemble of objects of a given mass. To approach this problem analytically, we apply the excursion set formalism (BCEK) that has been used to predict halo formation histories (LCa,b). Given a collapsed object of mass $M_2$ at some redshift $z_2$ this formalism predicts the probability distribution for the object to have a mass $M_1$ at some higher redshift $z_1$. We denote by $\sigma_M(M)$ the *rms* fluctuation amplitude $\delta M/M$ (using a $k$-space top-hat window function) as a function of mass $M$ and define $\delta_c(z)$ to be the overdensity required for the collapse of a spherical perturbation to occur at redshift $z$; both these quantities are measured at the present epoch as extrapolated in linear theory. Then the value of $M_1$ is found implicitly from the relation

$$\sigma_M^2(M_1) = \sigma_M^2(M_2) + \frac{[\delta_c(z_2) - \delta_c(z_1)]^2}{x^2}, \qquad (2)$$

where $x$ is a Gaussian-distributed random number with zero mean and unit variance (see eq. 2.16 of LCb and change variables to $x^2 = [\delta_{c1} - \delta_{c2}]^2/[\sigma_1^2 - \sigma_2^2]$). Using this equation, we start from a given mass at the present epoch and step backwards in time to find its mass at a sequence of higher redshifts. We continue this procedure until $M_1$ is less than 10% of the present-day mass. Note that because we start from the present-day mass of the object, we are restricting our sample to isolated galaxies. Because isolated spiral galaxies do satisfy the Tully-Fisher relation, this is an acceptable restriction.

Next, we need to convert a given mass history into a binding energy. We do so by assuming that the mass gained in an infinitesimal redshift interval was in fact accreted in the most regular fashion, i.e. in spherical symmetry and with a constant $dM/dz$. Any variation on these simplifying assumptions would introduce additional degrees of freedom, such as the orbital parameters or the internal structure of the merger components. For spherical accretion in an arbitrary cosmology, the infinitesimal energy added to the system is

$$dE \propto \frac{GM\,dM}{M^{1/3}} \left( \frac{5}{3g}\delta_c(z) - \Omega_R \right), \qquad (3)$$

where $\Omega_0$ is the present fraction of the critical density in non-relativistic matter, $\Omega_\Lambda$ is the present fraction of the critical density in the cosmological constant, $\Omega_R = 1 - \Omega_0 - \Omega_\Lambda$ at the present epoch, $M$ is the mass interior to the added shell, $dM$ is the shell mass, and

$$g = \frac{5}{2} \int_0^1 \frac{a^{3/2}\,da}{(\Omega_\Lambda a^3 + \Omega_R a + \Omega_0)^{3/2}}. \qquad (4)$$

The last factor in equation 3 is the generalization of the factor of $t_{\rm coll}^{2/3}$ in equation 1. For finite $\Delta z$ and $\Delta M$, we integrate over $\Delta M$ assuming $dM/dz = \Delta M/\Delta z$ to find $\Delta E$. We then sum over all the redshift steps to find the total energy, which we set proportional to the square of the velocity dispersion.

As discussed in LCa, the above merger histories allow the possibility that the mass accreted in a given time step ($\Delta M$) may be greater than the mass in the progenitor ($M_1$). This is in conflict with our goal to track the mass of the largest progenitor of an object. In the context of equation 3, it means that we are allowing most of the object to accrete spherically onto a small central seed in a short redshift interval. As this process seems unrealistic, we reverse the roles of the two pieces and consider a small spherical accretion onto a larger mass central piece. In each redshift step $\Delta z$, if the mass $M_1$ found from equation 2 is less than half of $M_2$, we flip the roles of the pieces and instead take the mass to be $M_2 - M_1$. LCa use a slightly different scheme to remedy this problem when constructing their Monte Carlo merger history algorithm. Surprisingly, when they compare the formalism to N-body simulations (LCb), the statistics of the time at which a present-day object has acquired half its mass better match the unconstrained formalism rather than the Monte Carlo formalism, despite the fact that following the more massive of the two merging pieces seems more appropriate. In this work, we use the constrained histories (i.e. tracking the more massive piece), since they produce about 15% less scatter in the $M$–$\sigma$ relation than the unconstrained histories.

Fig. 1.— Fractional scatter in the velocity dispersion $\sigma$ as a function of the present-day overdensity $\delta M/M \equiv \sigma_M(M_0)$ on the mass scale of the final object $M_0$. The different curves correspond to scale-free power spectra, $P(k) \propto k^n$, with different values of $n$, in an $\Omega = 1$ cosmology. The "No Mergers" case requires that the object never accrete more than 20% of its final mass in any given time step.

We may now integrate over many generated mass histories to find the variation in the velocity dispersion on a given mass scale. We first consider scale-free power-spectra $P(k) \propto k^n$, for which $\sigma_M \propto M^{-(n+3)/6}$, and find $\Delta\sigma/\sigma$ as a function of $\sigma_M(M_0)$ and $n$, where $M_0$ is the present-day mass and $\Delta\sigma$ is the standard deviation of the distribution of velocity dispersions $\sigma$. The results are shown in Figure 1 and compared to the scatter in luminosity that would result if we enforced a Tully-Fisher relation of the type $L \propto \sigma^3$ (dotted horizontal lines). The figure shows that the scatter in $\sigma$ drops strongly with decreasing $\sigma_M(M_0)$ and with more negative $n$. Both these changes tend to bring the collapse to lower redshift so that the ensemble of objects is described by a narrower range of collapse redshifts. An open universe ($\Omega_0 < 1$, $\Lambda = 0$) results in a larger scatter at a given $\sigma_M(M_0)$ and $n$. A flat universe with $\Lambda \neq 0$ produces the same scatter as its flat $\Lambda = 0$ counterpart, because if $\Omega_R = 0$ then both equations 2 and 3 involve only $\delta_c$ and functions of mass, with $\Lambda$ entering only in the constant of proportionality of the energy and therefore dropping out of $\Delta\sigma/\sigma$.

Fig. 2.— Fractional scatter in the velocity dispersion $\sigma$ in four different CDM models. The flat, low-$\Omega_0$ model is the only one with a non-zero cosmological constant. The parameter $\Gamma = \Omega_0 h$ is used to fit the power spectrum (cf. Efstathiou et al. 1992).

We next consider various cold dark matter (CDM) models, as parameterized by the constant $\Gamma$ (Efstathiou et al. 1992). We normalize the power-spectrum to COBE using $\sigma_8$, the *rms* amplitude of fluctuations in spheres of radius $8h^{-1}$ Mpc (Górski et al. 1995a,b; Stompor et al. 1995). Figure 2 presents the resulting scatter as a function of mass scale. We find that $\Omega = 1$, COBE-normalized CDM (solid line) produces too much scatter; this reflects the tendency of this model to have too much power on small-scales. Under-normalized $\Omega = 1$ CDM (dotted line) produces less scatter, while open and $\Lambda$-dominated universes give yet lower values. The lower value of $\sigma_8$ is the major factor in reducing the spread. However lower values of $\sigma_8$ than the ones shown in the figure are not allowed by the abundance of clusters of galaxies (White et al. 1993).

Finally, we consider the actual $M$–$\sigma$ relation rather than its expected scatter. If all of the objects had the same formation history regardless of mass scale, then $E \propto M\sigma^2 \propto M^{5/3}$, would yield $M \propto \sigma^3$. Instead we find a weak tendency for smaller objects (with higher $\sigma_M[M_0]$) to collapse at higher redshifts and therefore to have more binding energy. We therefore find $M \propto \sigma^\alpha$, with $\alpha \approx 3.1-3.2$. Although a weak dependence of $M/L$ on $M$ may lower the expected exponent by a few tenths, this relation is in reasonable agreement with the observed Tully-Fisher relation in the R and I bands (Mathewson et al. 1992) as long as its scatter is ignored.

## 3. Discussion

We next examine the implications of the predicted scatter in $M$–$\sigma$ for the Tully-Fisher relation. In attempting to rule out a primordial origin for this relation, we adopt the most optimistic approach for connecting $M$–$\sigma$ to $L$–$v_c$. We assume that the circular velocity of the baryonic disk is strictly determined by the velocity dispersion of its spherical isothermal halo and that the overall luminosity of the galaxy is tightly related to the mass of its halo. The latter relation might exist if all the baryons in the halo cooled onto the central disk and were turned into stars with a universal mass-to-light relation. We then treat the scatter in $\sigma$ on a given mass scale as a minimal source of intrinsic scatter in the Tully-Fisher relation.

How much scatter is allowed by the observations? The full observed scatter in the largest samples is $0.4 \pm 0.02$ magnitudes (Willick 1995a,b,c). Many different effects contribute to this error budget. Observational errors are usually estimated to be 0.15–0.30 mag (Strauss & Willick 1995), with a significant contribution coming from the measurement of the HI line widths. The ellipticity of the halo potential leads to error in measuring the circular velocity and estimating the inclination of the baryonic disk (Franx & de Zeeuw 1992); these effects have been measured in a sample of spirals and estimated to contribute a scatter of about 0.15 mag (Rix & Zaritsky 1995). The above effects alone reduce the remaining error budget to $\lesssim 0.3$ mag, which corresponds to about 10% scatter in the circular velocity. It seems likely that variations in the mass function of stars would lead to additional scatter, leaving less of the error budget available for intrinsic variations in $\sigma$. In addition, other studies of the Tully-Fisher relation quote lower values for the observed scatter (Pierce & Tully 1988; Schommer et al. 1993; Bernstein et al. 1994). In summary, $\Delta\sigma/\sigma = 10\%$ provides an upper limit to the intrinsic scatter in $\sigma$ at fixed $M$ that is allowed by observations.

Naively, if each halo had a single formation redshift $z_f$, then the above 10% limit would require that all halos of a given mass form within an unusually narrow redshift interval, $\Delta z_f/(1+z_f) \lesssim 20\%$. Indeed, Figure 2 indicates that reasonable cosmologies are at best marginally consistent with the above 10% limit. Smaller masses are predicted to have a larger scatter, matching a trend in the data (Federspiel, Sandage, & Tammann 1994), although various observational effects could also introduce larger scatter in dimmer galaxies.

In calculating the minimum Tully-Fisher scatter due to differences in the formation histories of dark halos, we have introduced two simplifying assumptions. First, we assumed spherical accretion in relating the mass history of an object to its final binding energy $E$ (cf. Eq. 3). Non-sphericity would add new degrees of freedom to our analysis, such as the orbital parameters or the internal structure of the merger components. These additional parameters could increase the resulting $M$–$\sigma$ scatter. Second, we have assumed that the mass and stellar luminosity of the gas that cools to form the galactic disk is a strict function of the mass of the halo. The ratio between the disk mass and the total mass of spiral galaxies is $\lesssim 5$–10% (Zaritzky & White 1994; Kochanek 1995). As this fraction is somewhat lower than the baryon fraction observed in clusters of galaxies (David, Jones, & Forman 1995), it is possible that not all of the baryons originally associated with a halo are currently in its stellar disk. If the fraction of baryons included in the disk varies from one galaxy to another, then the total luminosity of the galaxy will not be tightly constrained by the mass of its halo and the expected Tully-Fisher scatter would increase. As for the fate of the remaining baryons, they may either be expelled from the galaxy by supernova-driven winds (e.g. Spitzer 1956; Corbelli & Salpeter 1988) or remain in the halo in the form of hot pressure-supported gas or as dark compact objects. X-ray observations exclude the existence of extended x-ray halos or cooling flows in most spiral galaxies (Fabbiano 1989). Searches for microlensing events toward the LMC indicate that a noticeable fraction of the halo mass may consist of baryonic objects (Alcock et al. 1995; Gates, Gyuk, & Turner 1994). If these objects formed in the intergalactic medium at very high redshift, then they would be essentially indistinguishable from other types of cold dark matter. However, if they formed within proto-galaxies, then their mass fraction may vary from galaxy to galaxy. Similarly, if supernovae are effective at expelling gas out of the disk, then the amount of gas expelled would likely vary from one galaxy to another due to differences in their star formation history. Any such variations would contribute additional scatter to the relation.

The Tully-Fisher relation applies to a particular morphological class of galaxies, and so one may argue that galaxies which have undergone major mergers since the formation of their stellar disks would not appear as thin spiral galaxies today (Tóth & Ostriker 1992). This morphological restriction on merger histories ought to yield a smaller value of $\Delta\sigma/\sigma$, but because the epoch of star formation is unknown, it cannot be incorporated in a simple way. Nevertheless, in order to examine the significance of this restriction, we have modified our algorithm so that any object that has undergone a merger that added more than 20% of its present-day mass is rejected. This constraint rejected 70% of the systems and reduced the scatter in $v_c$ by a factor of 20% (cf. the lowest curve in Fig. 1). Setting the rejection threshold at 30% in mass rejected 30% of the systems and reduced the scatter by 10%. While this morphological constraint goes in the right direction, it leads to a small overall effect.

Another way to implement the above morphological restriction is to argue that spiral galaxies already exist in their current form at some high redshift, $z_f$. If accretion does not affect the rotation curve or luminosity of these galaxies between this early redshift and the present time, then the effective value of $\sigma_M(M_0)$ would be lowered by the ratio of the growth factors at $z_f$ and today, e.g. by $(1+z_f)$ in an $\Omega=1$ universe. Note that the fact that structure formation cuts off at late times in an open universe is not sufficient for this purpose. As Figures 1 and 2 indicate, lowering $\sigma_M(M_0)$ reduces the scatter in the $M$–$\sigma$ relation. At first glance, it seems unlikely that both the baryonic mass and the velocity dispersion of the halo would remain unchanged as the growth factor increases by a significant factor and the corresponding accretion occurs. However, detailed consideration of these conditions raises difficult questions concerning the maximum mass of galaxies (Thoul & Weinberg 1995), the accretion of satellite galaxies (Tóth & Ostriker 1992; Navarro, Frenk, & White 1994), and the fate of dark matter cores as they fall into larger objects (Moore, Katz, & Lake 1995). It is important to note that galaxy-sized halos in the universe at $z=0$ do not correspond to rare density peaks. Since $\sigma_M \approx 3$, they correspond to regions with a peak height $\nu \equiv \delta_c(0)/\sigma(M) \approx 0.5$. However, galaxy formation arguments have often associated galaxies with $\nu \approx 2$ peaks. The excursion set formalism predicts that if galaxies form out of $2\sigma$ peaks when the growth factor is one-third of its present value (i.e. $z=2$ in an $\Omega_0=1$ universe or $z=4.3$ in an open $\Omega_0=0.3$ universe), then 90% (50%) of these galaxies are incorporated today within systems at least 4.4 (37) times as massive. Here we assume an $n=-1$ power spectrum; more negative values of $n$ result in even larger systems. Hence, although galaxy formation at high redshift reduces the $M$–$\sigma$ scatter, one must explain why the systems are unchanged by their infall into larger objects.

Previous estimates of the $M$–$\sigma$ scatter caused by the properties of Gaussian random fields (Cole & Kaiser 1989) have assumed a Gaussian distribution of peak height $\nu$ and then used $\sigma \propto \nu^{1/2}$ to derive the scatter in $\sigma$. However, in the Press-Schechter formalism, objects of a particular mass at $z=0$ correspond to a single value of $\nu$. Peaks with larger $\nu$ formed earlier and are now part of more massive objects. Since the Tully-Fisher relation is measured at $z=0$, we cannot assign galaxies to a distribution of $\nu$. Instead, we must recognize that objects of a particular mass today were assembled through a variety of merger histories, leading to differences in their binding energies. It is interesting to note that these differences result in a smaller Tully-Fisher scatter ($\approx 0.4$ mag) than that inferred from arguments based on the distribution of $\nu$ ($\gtrsim 1$ mag).

A different approach to the Tully-Fisher relation relies on the notion that the cooling of the gas is incomplete (Kauffmann, White, & Guiderdoni 1993, hereafter KWG). Here, one assumes that after collapse the gas acquires an isothermal density profile similar to that of the dark matter. If one then considers the cooling time as a function of radius and includes in the galactic disk only the mass of the gas that has had time to cool, $M_{\rm cold}$, then the reservoir of cold gas depends on the velocity dispersion and not on the mass of the halo, since the former sets the density profile of the gas. Thus, it is possible to get a tight $\sigma - M_{\rm cold}$ relation regardless of the actual halo mass and its formation history. While this result is encouraging, there are some problems. The cooling of gas from the halo should be ongoing, with up to 10 $M_\odot$ yr$^{-1}$ of cold gas falling onto the disk, well in excess of the observed star formation rates (White 1994) and producing X-ray halos much brighter than observed (Fabbiano 1989). Even for a minimum steady-state accretion rate of a few $M_\odot$ yr$^{-1}$, necessary to accumulate a typical disk mass in a Hubble time, the dissipation of the radial-infall kinetic energy should already produce $\sim 10^{41}$ erg s$^{-1}$, well beyond the typical limits on diffuse X-ray emission from spirals, $\lesssim 5 \times 10^{39}$ erg s$^{-1}$ (Bregman & Glassgold 1982; Fabbiano 1989). The lack of X-ray halos in spiral galaxies suggests that gas accretion from the halo has essentially ended and that an isothermal gaseous halo cannot be the regulating mechanism for the Tully-Fisher relation. SPH simulations (Navarro & White 1994) support this view by showing that the halo gas tends to cool efficiently due to the clumping that characterizes hierarchical merging, so that rather little gas is left in X-ray halos at the present time. Although these simulations do not include heating from supernova-driven winds that might slow the cooling or return gas to the halo, it is unlikely that this heating will maintain the gas just at the virial temperature of the halo at all times.

Empirical evidence that cold gaseous disks have already condensed at early times comes from the fact that the mass of neutral hydrogen (HI) in damped Ly$\alpha$ systems at $z \approx 2 - 3$ is comparable to the total mass in stars of local disk galaxies (Lanzetta et al. 1995). The detected HI gas has a column density $\sim 10^{20-22}$ cm$^{-2}$, similar to that found in nearby galactic disks (Broeils & Van Woerden 1994). The damped Ly$\alpha$ absorption is often accompanied by abundant heavy-element absorption lines at low-ionization stages (Turnshek et al. 1989; Wolfe et al. 1993; Petini et al. 1994). The velocity field traced by these metal lines relative to the damped Ly$\alpha$ line is consistent with typical galactic rotation velocities, as if the damped component were associated with a galactic disk while the metal-line systems were embedded in a surrounding halo (Lanzetta & Bowen 1992; Lu et al. 1993; Turnshek & Bohlin 1993; Lu & Wolfe 1994). The association with galaxies is indeed confirmed by direct imaging in a few cases (Steidel et al. 1994, 1995). These facts suggest that a considerable fraction of the cold gas necessary to form the stellar content of galactic disks today was already available at $z \gtrsim 2$ and could not have been the result of a recent cooling process. The inferred metallicity of this high-redshift gas is an order of magnitude lower than solar (Pei, Fall, & Bechtold 1991; Sembach et al. 1995), implying subsequent star formation and metal enrichment. If the halos of disk galaxies were also formed at $z \gtrsim 1$ and have not accreted mass afterwards, then the low scatter in the Tully-Fisher relation could be explained.

Evrard, Summers, and Davis (1994) have performed an SPH simulation for an $\Omega = 1$ CDM model with $\sigma_8 = 0.6$. Many of the halos in their simulation had disks of cold gas at the center, and the correlation between the mass of the central disk and the circular velocity of the halo proved to have very little scatter, corresponding to ∼0.13–0.26 mag at a redshift of unity. However, since the analysis of the simulated galaxies was performed at $z = 1$, one should use $\sigma_8 = 0.3$ in comparing it to the $z = 0$ spherical model. At the mass scale considered, this leads to $\sigma_M = 1.8$ and $n = -2$, which, according to Figure 1, yields a scatter of 0.07 in $\sigma$ or about 0.2 mag in luminosity. Hence, the prediction of the spherical model is comparable to that of the simulation. The model also predicts that if the simulation were to be evolved down to $z = 0$, then the scatter in the estimated Tully-Fisher relation would have been substantially increased. Future simulations of this type would be particularly valuable as they could shed light on some of the issues raised above, including gas cooling, accretion in groups of galaxies, and non-sphericity.

We conclude that the expected scatter in the $M$–$\sigma$ properties of dark matter halos is too large to result in the observed Tully-Fisher relation. We see two paths for breaking the assumptions leading to this conclusion. First, if galaxies formed at redshifts $z \gtrsim 1$ and did not accrete material subsequently (contrary to expectations from the standard hierarchical clustering picture), then the scatter in the $M$–$\sigma$ relation would be significantly reduced. Second, if a strong feedback process comes into play, it may regularize star formation and gas dynamics and detach the overall luminosity of spiral galaxies from details concerning the formation history of their halos. Either one of these paths would have significant consequences for the process of galaxy formation. The similarity between the Faber-Jackson (1976) relation for ellipticals and the Tully-Fisher relation for spirals (cf. Peebles 1993) may hint that the actual path taken in reality is universal among different galaxy types.

We thank Tsafrir Kolatt, Jeffery Willick, Guiseppina Fabbiano, David Spergel, and David Weinberg for useful discussions. D.J.E. was supported in part by a National Science Foundation Graduate Research Fellowship.